\documentclass[sigconf]{acmart}
\usepackage{booktabs} % For formal tables

\usepackage{comment}
\usepackage{xcolor}
\usepackage{balance}

\copyrightyear{2022}
\acmYear{2022}
\setcopyright{rightsretained}

\acmConference[KDDCup '22]{To be presented at KDD Cup 2022 Workshop:
ESCI Challenge for Improving Product Search}{August 17, 2022}{Washington, DC, USA}
\acmBooktitle{KDD Cup 2022 Workshop: ESCI Challenge for Improving Product Search, August 17, 2022, Washington, DC, USA}
\acmDOI{}
\acmISBN{}

\makeatother

\begin{document}
\fancyhead{}
\title{%NeuralMind at Amazon KDD Cup 2022: Large Language Models for Product Search
A Boring-yet-effective Approach for the Product Ranking Task of the Amazon KDD Cup 2022}

\author{Vitor Jeronymo}
\affiliation{%
  \institution{NeuralMind, Brazil}
  \institution{University of Campinas, Brazil}
}

\author{Guilherme Rosa}
\affiliation{%
  \institution{NeuralMind, Brazil}
  \institution{University of Campinas, Brazil}
}

\author{Surya Kallumadi}
\affiliation{%
  \institution{Lowes, USA}
}
%\email{c@d.edu}

\author{Roberto Lotufo}
\affiliation{%
  \institution{NeuralMind, Brazil}
  \institution{University of Campinas, Brazil}
 }
%\email{c@d.edu}

\author{Rodrigo Nogueira}
\affiliation{%
  \institution{NeuralMind, Brazil}
  \institution{University of Campinas, Brazil}
}
%\email{c@d.edu}
%\renewcommand{\shortauthors}{A. SSSS, S. TTTT, T. KKKK}

\begin{abstract}
In this work we describe our submission to the product ranking task of the Amazon KDD Cup 2022. We rely on a receipt that showed to be effective in previous competitions: we focus our efforts towards efficiently training and deploying large language models, such as mT5, while reducing to a minimum the number of task-specific adaptations. Despite the simplicity of our approach, our best model was less than 0.004 nDCG@20 below the top submission. As the top 20 teams achieved an nDCG@20 close to 0.90, we argue that we need more difficult e-Commerce evaluation datasets to discriminate retrieval methods.

\end{abstract}




\settopmatter{printacmref=true}
\maketitle

\section{Introduction}

Recent improvements in information retrieval, mainly due to pretrained transformer models, opened up the possibility of improving search in various domains~\cite{Jin2020AliababaDA,macavaney2020sledge,bi2020transformer,choi2020semantic,lin2021pretrained,pradeep2021vera,nogueira2020navigation,ma2021retrieving}. Among such domains, e-commerce search receives special attention by the industry as improvements in search quality often lead to increases in revenue.

In this work, we detail our submission to the Amazon KDD Cup 2022, whose goal is to evaluate ranking methods that can be used to improve the customer experience when searching for products.

%However, only a handful of datasets for evaluating e-commerce search methods are publicly available~\cite{homedepot, wands, etc}, and often do not reflect ...

\section{Related Work}

Our solution is based on the monoT5 model, that demonstrated strong effectiveness in various passage ranking tasks in different domains. We qualify our method as ``boring'', since it is well known in the recent IR literature that models with more parameters can outperform smaller ones with task-specific adaptations. For example, \citet{nogueira2020document} used the model to achieve state-of-the-art results on TREC 2004 Robust Track \cite{trec2004} while \citet{pradeep2020h2oloo} used the same model, finetuned only on MS MARCO, to achieve the best or second best performance on medical domain ranking datasets, such as Precision Medicine~\cite{Roberts2019OverviewOT} and TREC-COVID~\cite{zhang2020rapidly}. In addition, Rosa et al. \cite{icail_2021, jurisin2022} used large versions of monoT5 to reach the state of the art in a legal domain entailment task in the COLIEE competition \cite{rabelo2021coliee, coliee2022summary}. Furthermore, \citet{rosa2022} showed that the 3 billion-parameter variant of monoT5 achieves the state of the art in 12 out of 18 datasets of the Benchmark-IR (BEIR) \cite{thakur2021beir}, which consists of datasets from different domains such as web, biomedical, scientific, financial and news.

\section{Methodology}

In this section, we describe mMonoT5, a multilingual variant of monoT5~\cite{nogueira2020document}, which is an adaptation of the T5 model \cite{raffel2020t5} for the passage ranking task. We first finetune a multilingual T5 model \cite{xue2021mt5} on the mMARCO dataset \cite{DBLP:journals/corr/abs-2108-13897}, which is the translated version of MS MARCO \cite{marco} in 9 languages. The model is trained to generate a ``yes'' or ``no'' token depending on the relevance of a document to a query. 

mMonoT5 uses the following input template:

\begin{equation}
\text{Query: } \hspace{0.1cm} q \ \ \text{ Document: }  \hspace{0.1cm} d \ \ \text{ Relevant:}
\end{equation}

\noindent where $q$ represents a query and $d$ represents a document that may or may not be relevant to the given query.

During inference, the model receives the same input prompt and estimates a score $s$ that quantifies the relevance of a document $d$ to a query $q$  by applying a softmax function to the logits of the "yes" and "no" tokens, and then taking the probability of the "yes" token as the final score. That is, 

\begin{equation}
s = P(\textrm{Relevant}=1 | d, q).
\end{equation}

After computing all scores for a given query, we rank then with respect to their scores. 

After finetuning on mMARCO, we further finetuned the model on the training data of tasks 1 and 2 of the competition.
We use the Beautiful Soup library to clean any remaining HTML tags that may appear in the product. Products are presented to the model as the concatenation of the fields product\_title, product\_description, product\_bullet\_point, product\_brand and product\_color\_name, joined by whitespaces.

During the competition we observed that using task 2 training data improved the model substantially. Hence, we used task 1 and 2 training data by transforming the labeled data classes to ``true'' if ‘exact’ and all other classes as ``false''. We use these tokens instead of ``yes'' and ``no'', used by the original mMonoT5.
We trained the model for 5 epochs, which takes about 72 hours in a TPU v3, using batches of 128 and maximum sequence length of 512 tokens.

\section{Results}

We show our results in Table~\ref{tab:results}.
Our best model achieved an nDCG@20 of 0.9012 and 0.9007 on the public and private test sets, respectively, placing us in the ninth place on the leaderboard and only 0.0036 behind the first position. 

Initially, we used the mMonoT5 base, with 580M parameters, finetuned on mMarco data to test the model's zero-shot capability. This model achieves an nDCG@20 of 0.864. Then we further finetuned it on the training data of the competition, which results in a nDCG@20 of 0.89, which later, the 3.7B parameter version surpassed by 0.0112 points.
We also tried translating the corpus and queries into English and using the monoT5-3B (English-only) finetuned on the competition data, but it could not out-do its multilingual counterpart.

\begin{table}[]
\centering%\centering\resizebox{0.5\textwidth}{!}{
\begin{tabular}{lcc}
\toprule
&  \multicolumn{2}{c}{\textbf{nDCG@20}} \\
\textbf{Model} & {\textbf{Public}} & {\textbf{Private}} \\
\midrule
monoT5-3B (dataset translated to En) & 0.8750 & - \\
mMonoT5-580M (mMARCO only) & 0.8640 & - \\
mMonoT5-580M &  0.8900 & - \\
mMonoT5-3.7B (our best submission) & 0.9012 & 0.9007 \\
\midrule
First place (team www) & 0.9057 & 0.9043\\
20th place (team we666) & 0.8933 & 0.8929 \\
\bottomrule
\end{tabular}
%}
\caption{Main results of the competition.}
\label{tab:results}
\end{table}

\section{Conclusion}

We described a boring but effective approach based on the multilingual variation of monoT5 that achieved competitive results in the product ranking task of the Amazon KDD Cup 2022.

%\nocite{ecom17,ecom18,ecom19,ecom20}
\bibliographystyle{ACM-Reference-Format}
\balance
\bibliography{references} 


\begin{thebibliography}{23}

%%% the Web address.

\ifx \showCODEN    \undefined \def \showCODEN     #1{\unskip}     \fi
\ifx \showDOI      \undefined \def \showDOI       #1{#1}\fi
\ifx \showISBNx    \undefined \def \showISBNx     #1{\unskip}     \fi
\ifx \showISBNxiii \undefined \def \showISBNxiii  #1{\unskip}     \fi
\ifx \showISSN     \undefined \def \showISSN      #1{\unskip}     \fi
\ifx \showLCCN     \undefined \def \showLCCN      #1{\unskip}     \fi
\ifx \shownote     \undefined \def \shownote      #1{#1}          \fi
\ifx \showarticletitle \undefined \def \showarticletitle #1{#1}   \fi
\ifx \showURL      \undefined \def \showURL       {\relax}        \fi
% invisible to TeX
\providecommand\bibfield[2]{#2}
\providecommand\bibinfo[2]{#2}
\providecommand\natexlab[1]{#1}
\providecommand\showeprint[2][]{arXiv:#2}

\bibitem[\protect\citeauthoryear{Bajaj, Campos, Craswell, Deng, Gao, Liu,
  Majumder, McNamara, Mitra, Nguyen, Rosenber, Song, Stoica, Tiwary, and
  Wang}{Bajaj et~al\mbox{.}}{2018}]%
        {marco}
\bibfield{author}{\bibinfo{person}{Payal Bajaj}, \bibinfo{person}{Daniel
  Campos}, \bibinfo{person}{Nick Craswell}, \bibinfo{person}{Li Deng},
  \bibinfo{person}{Jianfeng Gao}, \bibinfo{person}{Xiaodong Liu},
  \bibinfo{person}{Rangan Majumder}, \bibinfo{person}{Andrew McNamara},
  \bibinfo{person}{Bhaskar Mitra}, \bibinfo{person}{Tri Nguyen},
  \bibinfo{person}{Mir Rosenber}, \bibinfo{person}{Xia Song},
  \bibinfo{person}{Alina Stoica}, \bibinfo{person}{Saurabh Tiwary}, {and}
  \bibinfo{person}{Tong Wang}.} \bibinfo{year}{2018}\natexlab{}.
\newblock \showarticletitle{MS MARCO: A Human Generated MAchine Reading
  Comprehension Dataset}.
\newblock \bibinfo{journal}{{\em arXiv preprint arXiv:1611.09268\/}}
  (\bibinfo{year}{2018}).
\newblock


\bibitem[\protect\citeauthoryear{Bi, Ai, and Croft}{Bi et~al\mbox{.}}{2020}]%
        {bi2020transformer}
\bibfield{author}{\bibinfo{person}{Keping Bi}, \bibinfo{person}{Qingyao Ai},
  {and} \bibinfo{person}{W~Bruce Croft}.} \bibinfo{year}{2020}\natexlab{}.
\newblock \showarticletitle{A transformer-based embedding model for
  personalized product search}. In \bibinfo{booktitle}{{\em Proceedings of the
  43rd International ACM SIGIR Conference on Research and Development in
  Information Retrieval}}. \bibinfo{pages}{1521--1524}.
\newblock


\bibitem[\protect\citeauthoryear{Bonifacio, Campiotti, de~Alencar~Lotufo, and
  Nogueira}{Bonifacio et~al\mbox{.}}{2021}]%
        {DBLP:journals/corr/abs-2108-13897}
\bibfield{author}{\bibinfo{person}{Luiz~Henrique Bonifacio},
  \bibinfo{person}{Israel Campiotti}, \bibinfo{person}{Roberto de
  Alencar~Lotufo}, {and} \bibinfo{person}{Rodrigo Nogueira}.}
  \bibinfo{year}{2021}\natexlab{}.
\newblock \showarticletitle{mMARCO: {A} Multilingual Version of {MS} {MARCO}
  Passage Ranking Dataset}.
\newblock \bibinfo{journal}{{\em CoRR\/}}  \bibinfo{volume}{abs/2108.13897}
  (\bibinfo{year}{2021}).
\newblock
\showeprint{2108.13897}
\showURL{%
\url{https://arxiv.org/abs/2108.13897}}


\bibitem[\protect\citeauthoryear{Choi, Kallumadi, Mitra, Agichtein, and
  Javed}{Choi et~al\mbox{.}}{2020}]%
        {choi2020semantic}
\bibfield{author}{\bibinfo{person}{Jason~Ingyu Choi}, \bibinfo{person}{Surya
  Kallumadi}, \bibinfo{person}{Bhaskar Mitra}, \bibinfo{person}{Eugene
  Agichtein}, {and} \bibinfo{person}{Faizan Javed}.}
  \bibinfo{year}{2020}\natexlab{}.
\newblock \showarticletitle{Semantic product search for matching structured
  product catalogs in e-commerce}.
\newblock \bibinfo{journal}{{\em arXiv preprint arXiv:2008.08180\/}}
  (\bibinfo{year}{2020}).
\newblock


\bibitem[\protect\citeauthoryear{Jin, Tan, Chen, Yan, Huang, Zhang, and
  Liu}{Jin et~al\mbox{.}}{2020}]%
        {Jin2020AliababaDA}
\bibfield{author}{\bibinfo{person}{Qiao Jin}, \bibinfo{person}{Chuanqi Tan},
  \bibinfo{person}{Mosha Chen}, \bibinfo{person}{Ming Yan},
  \bibinfo{person}{Songfang Huang}, \bibinfo{person}{Ningyu Zhang}, {and}
  \bibinfo{person}{Xiaozhong Liu}.} \bibinfo{year}{2020}\natexlab{}.
\newblock \showarticletitle{Aliababa DAMO Academy at TREC Precision Medicine
  2020: State-of-the-art Evidence Retriever for Precision Medicine with
  Expert-in-the-loop Active Learning}. In \bibinfo{booktitle}{{\em TREC}}.
\newblock


\bibitem[\protect\citeauthoryear{Kim, Rabelo, Goebel, Yoshioka, Kano, and
  Satoh}{Kim et~al\mbox{.}}{2022}]%
        {coliee2022summary}
\bibfield{author}{\bibinfo{person}{Mi-Young Kim}, \bibinfo{person}{Juliano
  Rabelo}, \bibinfo{person}{Randy Goebel}, \bibinfo{person}{Masaharu Yoshioka},
  \bibinfo{person}{Yoshinobu Kano}, {and} \bibinfo{person}{Ken Satoh}.}
  \bibinfo{year}{2022}\natexlab{}.
\newblock \showarticletitle{COLIEE 2022 Summary: Methods for Legal Document
  Retrieval and Entailment}.
\newblock \bibinfo{journal}{{\em Proceedings of the Sixteenth International
  Workshop on Juris-informatics (JURISIN 2022)\/}} (\bibinfo{year}{2022}).
\newblock


\bibitem[\protect\citeauthoryear{Lin, Nogueira, and Yates}{Lin
  et~al\mbox{.}}{2021}]%
        {lin2021pretrained}
\bibfield{author}{\bibinfo{person}{Jimmy Lin}, \bibinfo{person}{Rodrigo
  Nogueira}, {and} \bibinfo{person}{Andrew Yates}.}
  \bibinfo{year}{2021}\natexlab{}.
\newblock \showarticletitle{Pretrained transformers for text ranking: Bert and
  beyond}.
\newblock \bibinfo{journal}{{\em Synthesis Lectures on Human Language
  Technologies\/}} \bibinfo{volume}{14}, \bibinfo{number}{4}
  (\bibinfo{year}{2021}), \bibinfo{pages}{1--325}.
\newblock


\bibitem[\protect\citeauthoryear{Ma, Shao, Liu, Liu, Zhang, and Ma}{Ma
  et~al\mbox{.}}{2021}]%
        {ma2021retrieving}
\bibfield{author}{\bibinfo{person}{Yixiao Ma}, \bibinfo{person}{Yunqiu Shao},
  \bibinfo{person}{Bulou Liu}, \bibinfo{person}{Yiqun Liu},
  \bibinfo{person}{Min Zhang}, {and} \bibinfo{person}{Shaoping Ma}.}
  \bibinfo{year}{2021}\natexlab{}.
\newblock \showarticletitle{Retrieving legal cases from a large-scale candidate
  corpus}.
\newblock \bibinfo{journal}{{\em Proceedings of the Eighth International
  Competition on Legal Information Extraction/Entailment, COLIEE2021\/}}
  (\bibinfo{year}{2021}).
\newblock


\bibitem[\protect\citeauthoryear{MacAvaney, Cohan, and Goharian}{MacAvaney
  et~al\mbox{.}}{2020}]%
        {macavaney2020sledge}
\bibfield{author}{\bibinfo{person}{Sean MacAvaney}, \bibinfo{person}{Arman
  Cohan}, {and} \bibinfo{person}{Nazli Goharian}.}
  \bibinfo{year}{2020}\natexlab{}.
\newblock \showarticletitle{SLEDGE-Z: A Zero-Shot Baseline for COVID-19
  Literature Search}. In \bibinfo{booktitle}{{\em Proceedings of the 2020
  Conference on Empirical Methods in Natural Language Processing (EMNLP)}}.
  \bibinfo{pages}{4171--4179}.
\newblock


\bibitem[\protect\citeauthoryear{Nogueira, Jiang, Cho, and Lin}{Nogueira
  et~al\mbox{.}}{2020a}]%
        {nogueira2020navigation}
\bibfield{author}{\bibinfo{person}{Rodrigo Nogueira}, \bibinfo{person}{Zhiying
  Jiang}, \bibinfo{person}{Kyunghyun Cho}, {and} \bibinfo{person}{Jimmy Lin}.}
  \bibinfo{year}{2020}\natexlab{a}.
\newblock \showarticletitle{Navigation-based candidate expansion and pretrained
  language models for citation recommendation}.
\newblock \bibinfo{journal}{{\em Scientometrics\/}} \bibinfo{volume}{125},
  \bibinfo{number}{3} (\bibinfo{year}{2020}), \bibinfo{pages}{3001--3016}.
\newblock


\bibitem[\protect\citeauthoryear{Nogueira, Jiang, Pradeep, and Lin}{Nogueira
  et~al\mbox{.}}{2020b}]%
        {nogueira2020document}
\bibfield{author}{\bibinfo{person}{Rodrigo Nogueira}, \bibinfo{person}{Zhiying
  Jiang}, \bibinfo{person}{Ronak Pradeep}, {and} \bibinfo{person}{Jimmy Lin}.}
  \bibinfo{year}{2020}\natexlab{b}.
\newblock \showarticletitle{Document Ranking with a Pretrained
  Sequence-to-Sequence Model}. In \bibinfo{booktitle}{{\em Proceedings of the
  2020 Conference on Empirical Methods in Natural Language Processing:
  Findings}}. \bibinfo{pages}{708--718}.
\newblock


\bibitem[\protect\citeauthoryear{Pradeep, Ma, Nogueira, and Lin}{Pradeep
  et~al\mbox{.}}{2021}]%
        {pradeep2021vera}
\bibfield{author}{\bibinfo{person}{Ronak Pradeep}, \bibinfo{person}{Xueguang
  Ma}, \bibinfo{person}{Rodrigo Nogueira}, {and} \bibinfo{person}{Jimmy Lin}.}
  \bibinfo{year}{2021}\natexlab{}.
\newblock \showarticletitle{Vera: Prediction techniques for reducing harmful
  misinformation in consumer health search}. In \bibinfo{booktitle}{{\em
  Proceedings of the 44th International ACM SIGIR Conference on Research and
  Development in Information Retrieval}}. \bibinfo{pages}{2066--2070}.
\newblock


\bibitem[\protect\citeauthoryear{Pradeep, Ma, Zhang, Cui, Xu, Nogueira, and
  Lin}{Pradeep et~al\mbox{.}}{2020}]%
        {pradeep2020h2oloo}
\bibfield{author}{\bibinfo{person}{Ronak Pradeep}, \bibinfo{person}{Xueguang
  Ma}, \bibinfo{person}{Xinyu Zhang}, \bibinfo{person}{Hang Cui},
  \bibinfo{person}{Ruizhou Xu}, \bibinfo{person}{Rodrigo Nogueira}, {and}
  \bibinfo{person}{Jimmy Lin}.} \bibinfo{year}{2020}\natexlab{}.
\newblock \showarticletitle{H2oloo at trec 2020: When all you got is a
  hammer... deep learning, health misinformation, and precision medicine}.
\newblock \bibinfo{journal}{{\em Corpus\/}} \bibinfo{volume}{5},
  \bibinfo{number}{d3} (\bibinfo{year}{2020}), \bibinfo{pages}{d2}.
\newblock


\bibitem[\protect\citeauthoryear{Rabelo, Goebel, Kim, Yoshioka, Kano, and
  Satoh}{Rabelo et~al\mbox{.}}{2021}]%
        {rabelo2021coliee}
\bibfield{author}{\bibinfo{person}{Juliano Rabelo}, \bibinfo{person}{Randy
  Goebel}, \bibinfo{person}{Mi-Young Kim}, \bibinfo{person}{Masaharu Yoshioka},
  \bibinfo{person}{Yoshinobu Kano}, {and} \bibinfo{person}{Ken Satoh}.}
  \bibinfo{year}{2021}\natexlab{}.
\newblock \showarticletitle{Summary of the Competition on Legal Information
  Extraction/Entailment (COLIEE) 2021}.
\newblock \bibinfo{journal}{{\em Proceedings of the Eighth International
  Competition on Legal Information Extraction/Entailment\/}}
  (\bibinfo{year}{2021}).
\newblock


\bibitem[\protect\citeauthoryear{Raffel, Shazeer, Roberts, Lee, Narang, Matena,
  Zhou, Li, and Liu}{Raffel et~al\mbox{.}}{2020}]%
        {raffel2020t5}
\bibfield{author}{\bibinfo{person}{Colin Raffel}, \bibinfo{person}{Noam
  Shazeer}, \bibinfo{person}{Adam Roberts}, \bibinfo{person}{Katherine Lee},
  \bibinfo{person}{Sharan Narang}, \bibinfo{person}{Michael Matena},
  \bibinfo{person}{Yanqi Zhou}, \bibinfo{person}{Wei Li}, {and}
  \bibinfo{person}{Peter~J. Liu}.} \bibinfo{year}{2020}\natexlab{}.
\newblock \showarticletitle{Exploring the Limits of Transfer Learning with a
  Unified Text-to-Text Transformer}.
\newblock \bibinfo{journal}{{\em Journal of Machine Learning Research\/}}
  \bibinfo{volume}{21}, \bibinfo{number}{140} (\bibinfo{year}{2020}),
  \bibinfo{pages}{1--67}.
\newblock
\showURL{%
\url{http://jmlr.org/papers/v21/20-074.html}}


\bibitem[\protect\citeauthoryear{Roberts, Demner-Fushman, Voorhees, Hersh,
  Bedrick, Lazar, and Pant}{Roberts et~al\mbox{.}}{2019}]%
        {Roberts2019OverviewOT}
\bibfield{author}{\bibinfo{person}{Kirk Roberts}, \bibinfo{person}{Dina
  Demner-Fushman}, \bibinfo{person}{E. Voorhees}, \bibinfo{person}{W. Hersh},
  \bibinfo{person}{Steven Bedrick}, \bibinfo{person}{Alexander~J. Lazar}, {and}
  \bibinfo{person}{S. Pant}.} \bibinfo{year}{2019}\natexlab{}.
\newblock \showarticletitle{Overview of the TREC 2019 Precision Medicine
  Track}.
\newblock \bibinfo{journal}{{\em The ... text REtrieval conference : TREC. Text
  REtrieval Conference\/}}  \bibinfo{volume}{26} (\bibinfo{year}{2019}).
\newblock


\bibitem[\protect\citeauthoryear{Rosa, Bonifacio, Jeronymo, Abonizio, Fadaee,
  Lotufo, and Nogueira}{Rosa et~al\mbox{.}}{2022a}]%
        {rosa2022}
\bibfield{author}{\bibinfo{person}{Guilherme~Moraes Rosa},
  \bibinfo{person}{Luiz Bonifacio}, \bibinfo{person}{Vitor Jeronymo},
  \bibinfo{person}{Hugo Abonizio}, \bibinfo{person}{Marzieh Fadaee},
  \bibinfo{person}{Roberto Lotufo}, {and} \bibinfo{person}{Rodrigo Nogueira}.}
  \bibinfo{year}{2022}\natexlab{a}.
\newblock \showarticletitle{No Parameter Left Behind: How Distillation and
  Model Size Affect Zero-Shot Retrieval}.
\newblock \bibinfo{journal}{{\em arXiv preprint arXiv:2206.02873\/}}
  (\bibinfo{year}{2022}).
\newblock


\bibitem[\protect\citeauthoryear{Rosa, Bonifacio, Jeronymo, Abonizio, Lotufo,
  and Nogueira}{Rosa et~al\mbox{.}}{2022b}]%
        {jurisin2022}
\bibfield{author}{\bibinfo{person}{Guilherme~Moraes Rosa},
  \bibinfo{person}{Luiz Bonifacio}, \bibinfo{person}{Vitor Jeronymo},
  \bibinfo{person}{Hugo Abonizio}, \bibinfo{person}{Roberto Lotufo}, {and}
  \bibinfo{person}{Rodrigo Nogueira}.} \bibinfo{year}{2022}\natexlab{b}.
\newblock \showarticletitle{Billions of Parameters Are Worth More Than
  In-domain Training Data: A case study in the Legal Case Entailment Task}.
\newblock \bibinfo{journal}{{\em arXiv preprint arXiv:2205.15172\/}}
  (\bibinfo{year}{2022}).
\newblock


\bibitem[\protect\citeauthoryear{Rosa, Rodrigues, Lotufo, and Nogueira}{Rosa
  et~al\mbox{.}}{2021}]%
        {icail_2021}
\bibfield{author}{\bibinfo{person}{Guilherme~Moraes Rosa},
  \bibinfo{person}{Ruan~Chaves Rodrigues}, \bibinfo{person}{Roberto Lotufo},
  {and} \bibinfo{person}{Rodrigo Nogueira}.} \bibinfo{year}{2021}\natexlab{}.
\newblock \showarticletitle{To Tune or Not To Tune? Zero-shot Models for Legal
  Case Entailment}.
\newblock \bibinfo{journal}{{\em ICAIL’21, Eighteenth International
  Conference on Artificial Intelligence and Law, June 21–25, 2021, São
  Paulo, Brazil\/}} (\bibinfo{year}{2021}).
\newblock


\bibitem[\protect\citeauthoryear{Thakur, Reimers, Rücklé, Srivastava, and
  Gurevych}{Thakur et~al\mbox{.}}{2021}]%
        {thakur2021beir}
\bibfield{author}{\bibinfo{person}{Nandan Thakur}, \bibinfo{person}{Nils
  Reimers}, \bibinfo{person}{Andreas Rücklé}, \bibinfo{person}{Abhishek
  Srivastava}, {and} \bibinfo{person}{Iryna Gurevych}.}
  \bibinfo{year}{2021}\natexlab{}.
\newblock \showarticletitle{BEIR: A Heterogenous Benchmark for Zero-shot
  Evaluation of Information Retrieval Models}.
\newblock \bibinfo{journal}{{\em arXiv preprint arXiv:2104.08663\/}}
  (\bibinfo{date}{4} \bibinfo{year}{2021}).
\newblock
\showURL{%
\url{https://arxiv.org/abs/2104.08663}}


\bibitem[\protect\citeauthoryear{Voorhees}{Voorhees}{2004}]%
        {trec2004}
\bibfield{author}{\bibinfo{person}{Ellen~M. Voorhees}.}
  \bibinfo{year}{2004}\natexlab{}.
\newblock \showarticletitle{Overview of the TREC 2004 Robust Track}.
\newblock \bibinfo{journal}{{\em Proceedings of the Thirteenth Text REtrieval
  Conference, TREC 2004, Gaithersburg, Maryland, November 16-19, 2004\/}}
  (\bibinfo{year}{2004}).
\newblock


\bibitem[\protect\citeauthoryear{Xue, Constant, Roberts, Kale, Al-Rfou,
  Siddhant, Barua, and Raffel}{Xue et~al\mbox{.}}{2021}]%
        {xue2021mt5}
\bibfield{author}{\bibinfo{person}{Linting Xue}, \bibinfo{person}{Noah
  Constant}, \bibinfo{person}{Adam Roberts}, \bibinfo{person}{Mihir Kale},
  \bibinfo{person}{Rami Al-Rfou}, \bibinfo{person}{Aditya Siddhant},
  \bibinfo{person}{Aditya Barua}, {and} \bibinfo{person}{Colin Raffel}.}
  \bibinfo{year}{2021}\natexlab{}.
\newblock \bibinfo{title}{mT5: A massively multilingual pre-trained
  text-to-text transformer}.
\newblock   (\bibinfo{year}{2021}).
\newblock
\showeprint[arxiv]{cs.CL/2010.11934}


\bibitem[\protect\citeauthoryear{Zhang, Gupta, Nogueira, Cho, and Lin}{Zhang
  et~al\mbox{.}}{2020}]%
        {zhang2020rapidly}
\bibfield{author}{\bibinfo{person}{Edwin Zhang}, \bibinfo{person}{Nikhil
  Gupta}, \bibinfo{person}{Rodrigo Nogueira}, \bibinfo{person}{Kyunghyun Cho},
  {and} \bibinfo{person}{Jimmy Lin}.} \bibinfo{year}{2020}\natexlab{}.
\newblock \showarticletitle{Rapidly Deploying a Neural Search Engine for the
  COVID-19 Open Research Dataset}. In \bibinfo{booktitle}{{\em Proceedings of
  the 1st Workshop on NLP for COVID-19 at ACL 2020}}.
\newblock


\end{thebibliography}

\end{document}